\documentstyle[aps]{revtex}
\begin{document}
\draft
\title{Comment on ``Pulsar Velocities and Neutrino Oscillations''}
\author{Yong-Zhong Qian}
\address{Physics Department, 161-33, California Institute of Technology,
	 Pasadena, California 91125}
\date{\today}
\maketitle
\begin{abstract}
In a recent Letter, Kusenko and Segr\`e proposed a new mechanism
to explain the observed proper motions of pulsars. Their
mechanism was based on the asymmetric neutrino emission induced
by neutrino oscillations in the protoneutron star magnetic field.
In this note
I point out that their estimate of the asymmetry in the neutrino
emission is incorrect. A proper calculation shows that their
mechanism at least requires a magnetic field of strength 
$B\sim 10^{16}$ G
in order to produce the observed average pulsar velocity.
\end{abstract}


In a recent Letter \cite{KS}, Kusenko and Segr\`e proposed a new
mechanism to explain the observed proper motions of pulsars.
According to these authors, 
the average pulsar ``kick'' velocity may be produced by
the anisotropy in the $\nu_\tau$ emission
resulting from the $\nu_e\rightleftharpoons\nu_\tau$ conversions
inside the protoneutron star with 
a magnetic field of strength
$B\sim 3\times 10^{14}$ G. However, these authors incorrectly
calculated the asymmetry $\Delta k/k$
in the third component of
the neutrino momentum 
(hereafter, some of their notation
is adopted).
In deriving $\Delta k/k$, they improperly assumed
$dN_e/dT=(\partial N_e/\partial T)_{\mu_e}$.
A proper calculation 
gives a much smaller $\Delta k/k$
than their estimate.

The $\nu_\tau$ energy flux at the
distorted effective $\tau$
neutrinosphere (defined by $r=r_0+\delta\cos\phi$) is
$F_{\nu_\tau}(r)\propto T^4$.
The temperature and electron number density structure near
$r_0$ can be characterized by the scale heights
$h_T=|d\ln T/dr|^{-1}_{r_0}$
and $h_{N_e}=|d\ln N_e/dr|^{-1}_{r_0}$, respectively. 
The electron number density $N_e$ 
is proportional to the electron
fraction $Y_e$ and the matter density $\rho$.
Numerical supernova
calculations \cite{burrows1,burrows2} show that 
$Y_e$ decreases with increasing radius below the $e$
neutrinosphere. Furthermore, 
$\rho$ decreases with increasing radius
much more steeply than $T$ does
inside the protoneutron
star \cite{burrows1}. Therefore, we have 
$h_{N_e}<|d\ln\rho/dr|^{-1}_{r_0}=h_\rho<h_T$.
The distortion of the effective $\tau$ neutrinosphere
can be measured by the dimensionless parameter
\[\alpha\equiv{\delta\over h_{N_e}}
={3\over 2}{eB\over\mu_e^2}
\approx 0.22
\left({20\ {\rm MeV}\over\mu_e}\right)^2
\left({B\over 10^{16}\ {\rm G}}\right).\]
In the region above the effective $\tau$ neutrinosphere
but below the $e$ neutrinosphere, the conditions
$Y_e\approx 0.1$ and
$\rho\approx 10^{12}$ g cm$^{-3}$ give
$\mu_e\approx$ 24 MeV. So we have $\delta\ll h_{N_e}$ 
for $B\lesssim 10^{16}$ G.
To the leading order in $\alpha$,
the asymmetry in the third component of the neutrino
momentum is 
\[{\Delta k\over k}\approx{1\over 6}
{\int_0^\pi F_{\nu_\tau}\cos\phi\sin\phi d\phi\over
\int_0^\pi F_{\nu_\tau}\sin\phi d\phi}
\approx {2\over 9}\alpha{h_{N_e}\over h_T}.\]
The next order contribution to $\Delta k/k$ depends on
$\alpha^3$. 

A $\Delta k/k$ of a few percent is required to explain
the observed average pulsar kick
velocity. 
For $h_{N_e}<h_\rho=h_T/3$ corresponding to $\rho\propto T^3$
\cite{burrows1},
the mechanism proposed in Ref. \cite{KS} at least requires
a magnetic field of strength $B\sim 10^{16}$ G
in order to produce $\Delta k/k\sim 0.01$.

I thank George Fuller, Dong Lai, and Petr Vogel
for helpful discussions.
This work was supported by a grant at Caltech.


\begin{references}
\bibitem{KS}
A. Kusenko and G. Segr\`e, Phys. Rev. Lett. {\bf 77}, 4872 (1996).
\bibitem{burrows1}
A. Burrows and T. J. Mazurek, Astrophys. J. {\bf 259}, 330 (1982).
\bibitem{burrows2}
A. Burrows and J. M. Lattimer, Astrophys. J. {\bf 307}, 178 (1986).
\end{references}
\end{document}